\documentclass[aps,showpacs]{revtex4}%
\usepackage{amsfonts}%
\usepackage{amsmath}%
\usepackage{amssymb}%
\usepackage{graphicx}

\begin{document}
\title{Nonlocal Double-Slit Interference with Pseudothermal Light}
\author{Lu Gao, Jun Xiong, Lu-Fang Lin, Wei Wang, Su-Heng Zhang, and Kaige Wang\footnote{Author to whom
correspondence should be addressed. Electronic address:
wangkg@bnu.edu.cn}}

\affiliation{Department of Physics, Applied Optics Beijing Area
Major Laboratory, Beijing Normal University, Beijing 100875,
China}
\begin{abstract}
We perform a nonlocal double-slit interference experiment with
pseudothermal light. The experimental result exhibits a typical
double-slit interference fringe in the intensity correlation
measurement, in agreement with the theoretical analysis by means
of the property of the second-order spatial correlation of field.

\end{abstract}
 \maketitle

Young's double-slit experiment provided crucial evidence for the wave nature
of light. The light from the two slits fell onto a screen and produced a
visible pattern of light and dark parallel bands called fringes. These light
and dark bands are due to constructive and destructive interference of the
light coming from each slit. To create a stable interference pattern, the
source of waves must be coherent in both time and space. Recently, Fonseca
et al\cite{monken99} reported a nonlocal double-slit interference experiment
using entangled photon pairs. The signal and idler photons generated by
spontaneous parametric down-conversion (SPDC) are scattered by two spatially
separated apertures: none of them is a double-slit but their superposition
at the same place forms a double-slit. The experimental result showed that
an interference fringe appeared in the two-photon coincidence measurement
whereas the individual signal and idler intensity profiles did not exhibit
any fringe. The effect was attributed to the nonlocal nature of quantum
interference.

Recent studies have shown that a thermal light source can play a
role similar to that of a two-photon entangled source in ''ghost''
imaging, ''ghost'' interference and subwavelength
interference\cite{lugiato04,cheng04,cao05,lugiato05,wang04,cai04,xiong05,zhai05,shih06,bor}.
In this paper, we report a nonlocal double-slit experiment using a
pseudothermal light source. Though the source in our experiment is
incoherent and does not exist any quantum entanglement, the
interference effect can still be carried out by such a nonlocal
double-slit.

The experimental setup shown in Fig. 1 is similar to that in Ref.
[1] with the exception that a pseudo-thermal light source replaces
the entangled two-photon source. The pseudothermal source is
obtained by passing a focused semiconductor laser beam of wavelength
660 nm through a slowly rotating (0.002 Hz) ground glass disk G.
P$_{1}$ and P$_{2}$ are a polarizer and a Glan prism, respectively,
which are used to modulate the light intensity. The quasi-thermal
light is separated by a 50/50 non-polarizing beamsplitter BS, which
is 3.4 cm distant from the ground glass G. A$_{1}$ and A$_{2}$ are
apertures whose superposition forms a double slit with the slit
width 250 $\mu $m and the distance between two slit centers 670 $\mu
$m. A$_{1}$ and A$_{2}$ are placed at the equal distance 4.7 cm from
BS. D$_{1}$ and D$_{2}$ are charged-coupled-device (CCD) located at
the same distance 85.3 cm from BS, and the two CCDs register the
intensity distributions $I_{1}(x_{1})$ and $I_{2}(x_{2})$ across the
beams.

Figure 2 shows the experimental results. Each intensity profile registered
by the two CCDs does not exhibit any interference-diffraction pattern. It is
clear that the source in the experiment is incoherent and random in
transverse direction. Then we count the normalized intensity correlation
between the two CCDs $g^{(2)}(x_{1},x_{2})=\langle
I_{1}(x_{1})I_{2}(x_{2})\rangle /(\langle I_{1}(x_{1})\rangle \langle
I_{2}(x_{2})\rangle )$ where one CCD scans the position while the other
detects the intensity at a fixed position $x=0$. We can see that the two
intensity correlations, $g^{(2)}(0,x_{2})$ and $g^{(2)}(x_{1},0),$ exhibit
the double-slit interference fringes though there is no real double-slit in
each arm.

The experimental results above can be explained by considering the spatial
correlation properties of the thermal light. When a thermal light beam is
divided into two beams, the spatial intensity correlation between them can
be written as
\begin{eqnarray}
\left\langle I_{1}(x_{1})I_{2}(x_{2})\right\rangle  &=&\langle E_{1}^{\ast
}(x_{1})E_{2}^{\ast }(x_{2})E_{2}(x_{2})E_{1}(x_{1})\rangle   \nonumber \\
&=&\langle I_{1}(x_{1})\rangle \langle I_{2}(x_{2})\rangle +|\langle
E_{1}^{\ast }(x_{1})E_{2}(x_{2})\rangle |^{2},  \label{4}
\end{eqnarray}%
\label{6}
\begin{eqnarray}
\left\langle I_{j}(x)\right\rangle  &=&(1/\sqrt{2\pi })\int h_{j}^{\ast
}(x,x_{0}^{\prime })h_{j}(x,x_{0})\widetilde{W}(x_{0}^{\prime
}-x_{0})dx_{0}^{\prime }dx_{0},\qquad (j=1,2)  \label{6a} \\
\left\langle E_{1}^{\ast }(x_{1})E_{2}(x_{2})\right\rangle  &=&(1/\sqrt{2\pi
})\int h_{1}^{\ast }(x_{1},x_{0}^{\prime })h_{2}(x_{2},x_{0})\widetilde{W}%
(x_{0}^{\prime }-x_{0})dx_{0}^{\prime }dx_{0},  \label{6b}
\end{eqnarray}%
where $h_{j}(x,x_{0})$ $(j=1,2)$ is the transfer function describing the
field propagation in each beam, and $\widetilde{W}(x^{\prime }-x)$ is the
first-order spatial correlation for the thermal light source.

For simplicity we consider the broadband limit for the source, i.e. $%
\widetilde{W}(x^{\prime }-x)\rightarrow \sqrt{2\pi }W_{0}\delta (x^{\prime
}-x)$, and the symmetric arrangement of both apertures and detectors in two
arms as showed in the experiment. Let $A_{1}(x)$ and $A_{2}(x)$ be
transmission functions of the apertures in the two arms, their superposition
forms a double-slit function
\begin{equation}
D(x)=A_{1}(x)A_{2}(x)=\left\{
\begin{array}{cc}
1 & \qquad (d-b)/2\leq |x|\leq (d+b)/2 \\
0 & \text{others}%
\end{array}%
\right. ,  \label{3}
\end{equation}%
where $b$ and $d$ are the slit width and the distance between the centers of
two slits, respectively. The transfer function in each arm is given by
\begin{equation}
h_{j}(x,x_{0})=\frac{k}{2\pi i\sqrt{L_{0}L}}\exp [ik(L_{0}+L)]\int
dx^{\prime }A_{j}(x^{\prime })\exp \left[ ik\left( \frac{(x^{\prime }-x)^{2}%
}{2L}+\frac{(x^{\prime }-x_{0})^{2}}{2L_{0}}\right) \right] ,\qquad (j=1,2)
\label{7}
\end{equation}%
where $L_{0}$ is the distance between the beamsplitter and the
aperture, and $L$ the distance between the aperture and the
detector. $k$ is the wave number of the beam. Thus we can
calculate
\label{5}
\begin{eqnarray}
\left\langle I_{j}(x)\right\rangle  &=&\frac{W_{0}k}{2\pi L}\int
A_{j}^{2}\left( x^{\prime }\right) dx^{\prime },\qquad (j=1,2)  \label{5a} \\
\left\langle E_{1}^{\ast }(x_{1})E_{2}(x_{2})\right\rangle  &=&\frac{W_{0}k}{%
\sqrt{2\pi }L}\exp \left[ i\frac{k}{2L}\left( x_{2}^{2}-x_{1}^{2}\right) %
\right] \widetilde{D}\left[ \frac{k}{L}\left( x_{1}-x_{2}\right) \right] ,
\label{5b}
\end{eqnarray}%
where $\widetilde{D}(q)=(2b/\sqrt{2\pi })$sinc$(qb/2)\cos (qd/2)$ is the
Fourier transform of the double-slit function $D(x).$ The intensity of each
beam in Eq. (\ref{5a}) is independent of the transverse position due to the
incoherence of the source, and in the intensity correlation (\ref{4}) it
gives rise to a background ( the first term in Eq. (\ref{4})). However, the
second term of Eq. (\ref{4}) containing coherent information is now $%
\left\vert \left\langle E_{1}^{\ast }(x_{1})E_{2}(x_{2})\right\rangle
\right\vert ^{2}\propto $sinc$^{2}\left[ \frac{bk}{2L}\left(
x_{1}-x_{2}\right) \right] \cos ^{2}\left[ \frac{dk}{2L}\left(
x_{1}-x_{2}\right) \right] $. Therefore the interference fringe can be
obtained in the intensity correlation measurement by scanning position in
one beam and fixing a position in the other beam. The theoretical curves in
Fig. 2 are in a good agreement with the experimental observation.

In summary, we have shown that the nonlocal double-slit interference
can be realized with pseudothermal light. Our theoretical analysis
demonstrated that two spatially separated apertures are correlated
in the second-order field correlation. When the two apertures are
placed at an equal distance from the beamsplitter, a typical
double-slit interference pattern is obtained in the intensity
correlation. Physically, this can be understood since, for the
thermal light, lensless imaging can occur at the symmetric position
of object with respect to the beamsplitter\cite{cao05} and it causes
a equivalent double-slit due to the superposition of the two
apertures. We can conclude that the nonlocal double-slit
interference effect should be attributed to the second-order spatial
correlation of the field.

This work was supported by the National Fundamental Research Program of
China Project No. 2006CB921404, and the National Natural Science Foundation
of China, Project No. 10574015.

Figure Captions:

FIG. 1. Sketch of the experiment setup.

FIG. 2. Experimental data of intensity distributions (triangles) and
normalized intensity correlations (circles) where D$_{1}$ detects
the intensity at a fixed position while D$_{2}$ registers the
intensity distribution across the beam in (a) and vice versa in (b).
Statistical averages are taken over 5000 CCD\ frames. Numerical
simulation of a typical double-slit interference fringe is shown by
solid lines.

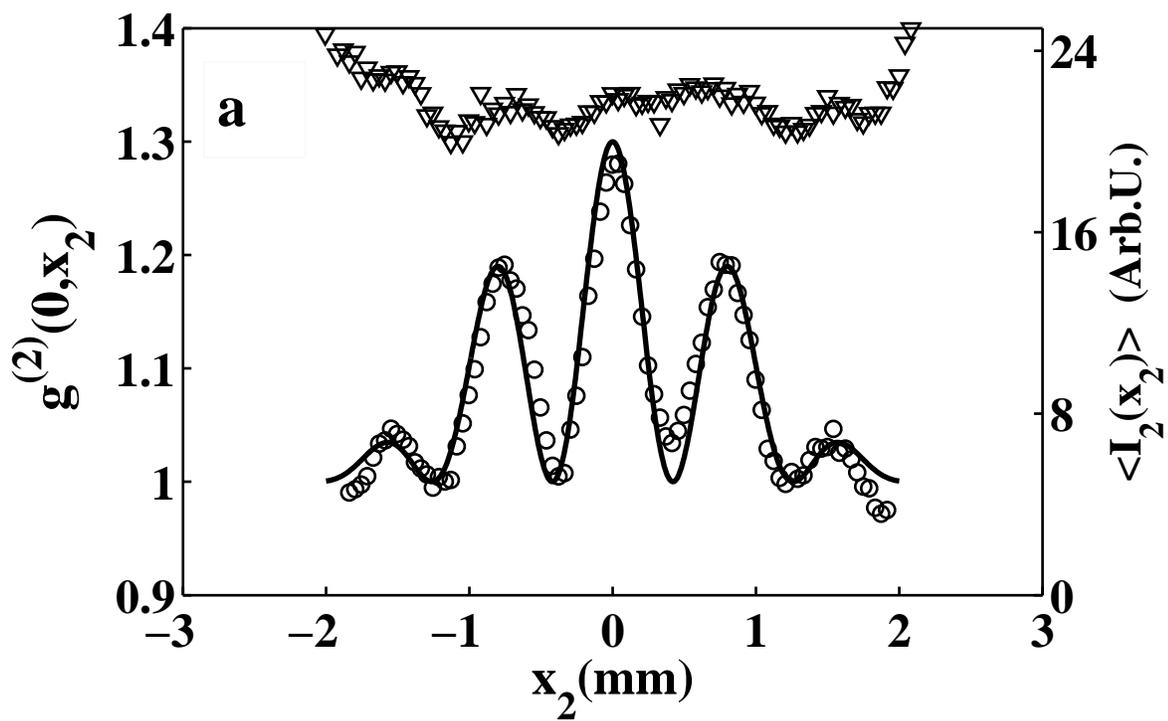
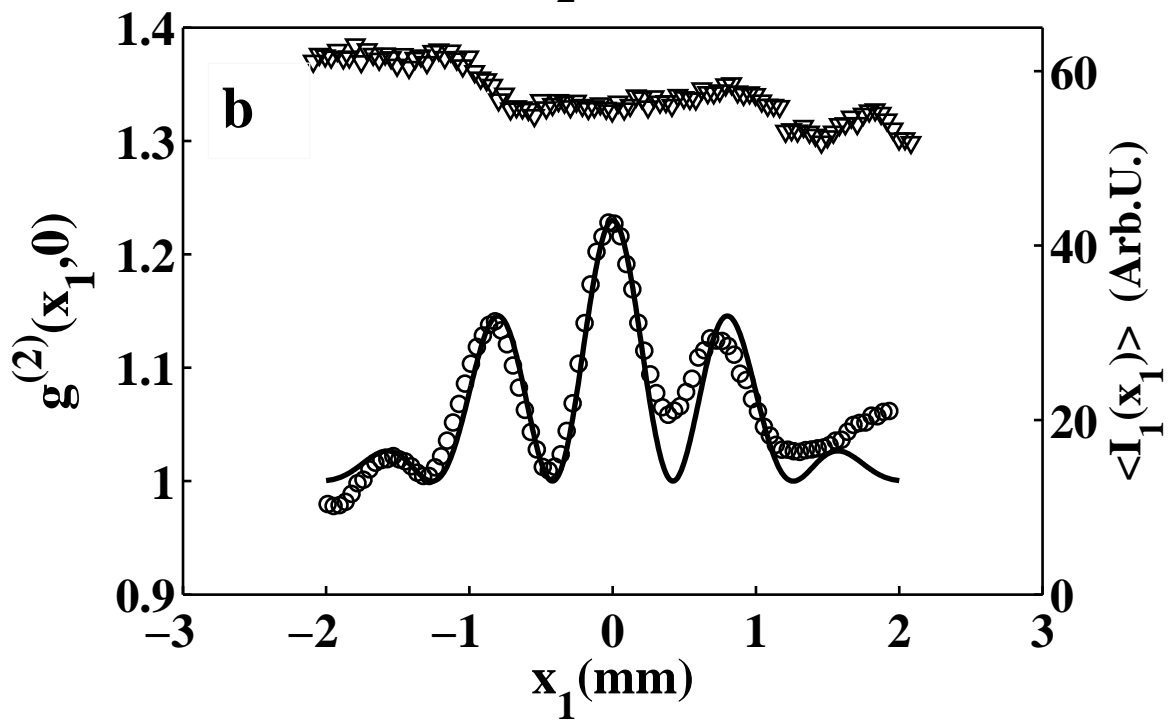
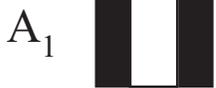
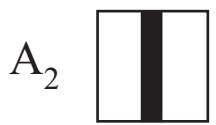
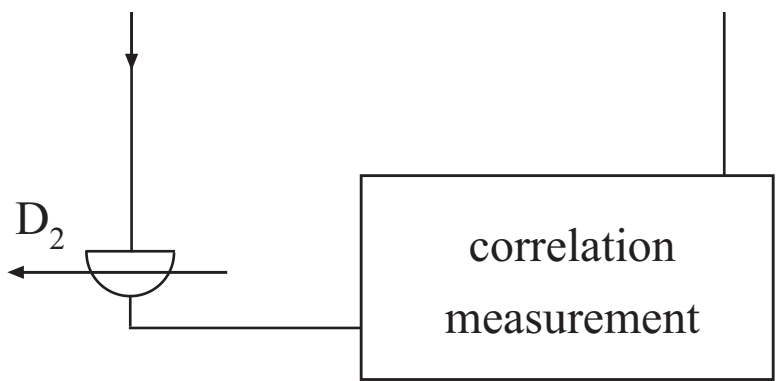
\end{document}